# Institutional Review Boards as Soft Governance Mechanisms of R&D:

Governing the R&D of AI-based Medical Products

*Antoni Lorente\**

*Risk-based approaches to governance bear an ambiguous stance regarding the R&D stages of AI, for they the possibility of explicit risks before they are posed by a given finalised product. In this context, Institutional Review Boards (IRBs) stand as unique governance mechanisms, capable of addressing the step from general research to concrete product development. However, IRBs face several challenges in governing AI-based medical products, including: (a) achieving consistency, (b) being exhaustive, (c) ensuring process transparency, and (d) reducing the existing capacity and knowledge asymmetry between different stakeholders. This article explores four governance levers that can be used to effect change, four governance entry-points throughout a product's lifecycle, and five different behaviours that IRBs should try to advance to ensure the effective governance of the R&D stages of AI-based medical projects. In doing so, IRBs can seize the unique opportunity they present to bring principles into practice, increase research quality, reduce governance costs, and bridge the knowledge gap between stakeholders.*

## I. Introduction

Institutional Review Boards (IRBs) have played a critical role in providing a channel to translate research ethics standards (such as The Belmont Report [1978] or the Declaration of Helsinki [1964]) into concrete practices, overseeing research proposals in healthcare for over 6 decades.[1] However, the efforts to include Artificial Intelligence (AI) across all sectors is pushing the limits of competence of IRBs when it comes to evaluating AI-based medical products, presenting unique challenges and opportunities for these boards to effectively govern the development of AI.

By considering the lifecycle of AI research projects and the current regulatory context (which is heavily leaning towards risk-based approaches), this article discusses IRBs as unique governance mechanisms to tackle AI at the research and development (R&D) stages – that is, the process of developing algorithms and architectures, training models, and developing them into concrete products or services.

Over the last years, and with AI becoming a promising avenue in healthcare research, IRBs' mandate has been expanded to incorporate AI-enabled medical products and research proposals as part of their scope. But the rapid development within the field of AI is imposing a heavy burden on IRBs that manifests via different challenges these boards face – of which this article considers four: (a) achieving consistency, (b) being exhaustive, (c) ensuring process transparency, and (d) addressing the existing capacity and knowledge asymmetry between certain private actors leading the development of AI and some independent evaluators. Failing to address these challenges may not only result in IRBs playing

---

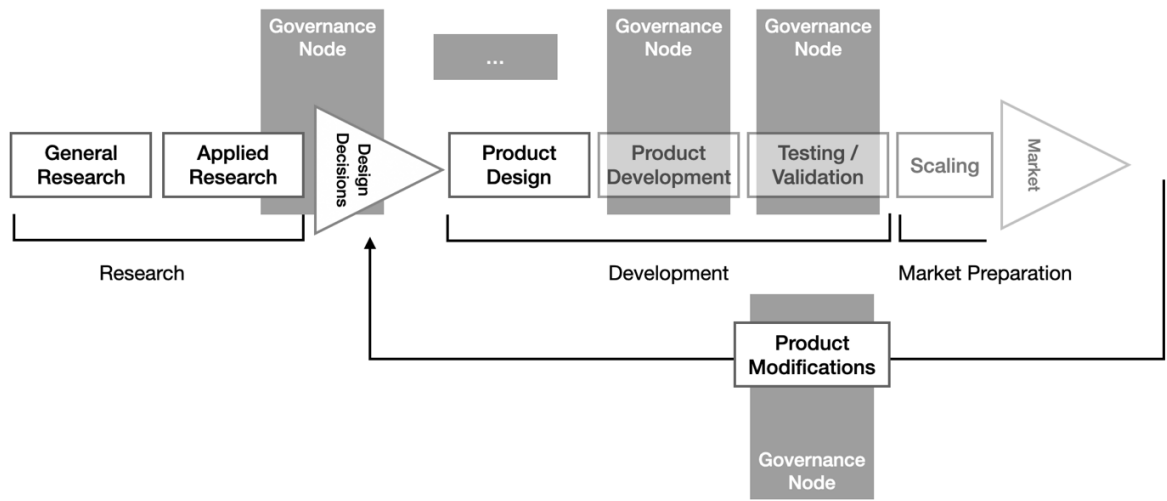

*Figure 1.* High-level representation of the product development pipeline to illustrate key stages of R&D, including the design decisions that divide research from development stages and different governance nodes

a role for which they have not been given the tools and resources to fulfill but could also mean the loss of a unique governance opportunity for AI research in healthcare (given that, in most jurisdictions, the risks and dangers stemming from AI systems tend to be defined after systems are deployed).

To address these challenges, this article develops in 8 stages. Section 1 provides a brief introduction. Section II presents IRBs and their broad regulatory context in the EU and the US. Section III presents the four main challenges that IRBs face to remain effective in governing the safe development of AI-enabled medical products. Section IV discusses IRBs in the AI context as critical governance mechanisms due to their pre-deployment scope. Section V presents four different governance levers or paths and mechanisms to ensure effective governance (ie, monitoring, harmonising, observing norms and standards and training) and their sources of authority. Section VI presents four governance nodes or entry-points throughout a product's lifecycle (ie, the steps from general research to proposal drafting, from project proposal to product development, results publication, and modifications or updates). Section VII identifies five key behaviours that different actors should include to empower and equip IRBs. Finally, Section VIII concludes with a final discussion on opportunities and limitations.

## II. Problem and Context

During the 1980s and 1990s, some of the world's leading economic powers went through a de-regulatory process that was, in part, triggered by a political move seeking to reduce the burden and overall costs of regulation industry. Across the UK, Europe, and US, different political waves tipped the scales towards a leaner, evidence-based, and industry-inspired regulatory approach. In this context, the private sector became the benchmark and gold standard for new regulatory proposals, percolating cost-benefit analyses, 'objective and transparent' and, ultimately, risk-based approaches into the public sector.[2]

Since then, the power of the 'central government' has been fragmented and crystallised into industry-specific or independent regulatory agencies, and alternative forms of self-regulation have been encouraged.[3] In parallel, succeeding industry-led technological booms throughout the last decades have reinforced this multi-channel approach to governance,

---

2   Bridget M. Hutter, 'The attractions of risk-based regulation: accounting for the emergence of risk ideas in regulation' (2005) 33 London: CARR <https://web.actuaries.ie/sites/default/files/erm-resources/205_The_attractions_of_risk_based_regulation_accounting_for_the_emergence_of_risk_ideas_in_regulation.pdf> accessed 12 August 2023.

3   Ibid.

where different mechanisms and stakeholders control different risks entailed by a given technology.[4]

When the concept of 'risk' was borrowed and industry adopted it as the backbone of some new approaches to governance, the lack of a neatly defined notion of risk in the legal context gave rise to several questions — most of which remain unanswered. In this sense, and in the European context, for example, regulatory crutches like tort law or product liability directives may have to cover the gap left by how the criticality of the risks related to AI are divided.[5] This is particularly important considering how the efforts by the industry to govern technological progress has prioritised operational risks or hazards faced by the industry over those faced by other stakeholders.

Recent and upcoming regulatory proposals in the European Union, such as the AI Act or the AI Liability Directive, are intended to address the new risks that new technologies like AI introduce.[6] On the one hand, the AI Act proposes a harmonised set of rules to ensure the safe and trustworthy development, deployment, and use of AI in the EU. On the other hand, and for reasons ranging from the information asymmetry between the regulator and AI developers to the complexity of the AI value chain, the AI Liability Directive is intended to 'protect consumers' liability claims for damage caused by AI-enabled products and services'.[7]

When it comes to EU's AI-healthcare applications, efforts have been mostly directed towards developing an infrastructure capable of supporting forthcoming requirements to evaluate AI-enabled medical products. In this sense, both the European Medicines Agency (EMA) and the European Commission (EC) have published different reports[8,9] laying out the priorities and objectives for Europe. However, all medical devices – including those enabled by AI – are subject to and must, comply with existing EU regulation.[10] In practice, IRBs have been stretched to include and evaluate AI-based research projects without necessarily having the skillset to do so, or without having a clear mandate and strict protocols to address these types of projects.

Moreover, the EU has notable variation in terms of regulation, applicable laws, and other relevant practices among different countries, which can sometimes lead to a lack of clarity for international projects (illustrating the coherence and consistency challenges discussed in Section 3) and, in some cases, to inefficiencies, highlighting the need for greater procedural harmonisation not only to reduce costs but also to increase the quality of the evaluation and prevent resource waste.[11]

In the US, the FDA is responsible for regulating IRB's and how they address AI research projects in healthcare.[12] In 2017, the FDA adopted the principles governing software as medical device (SaMD) developed by the International Medical Device Regulators Forum (IMDRF), a voluntary group of medical device regulators. These were later developed into several Food and Drug Administration (FDA) texts which covered, in part, the safe and effective development of AI-based software.[13] As a result, AI-based software developed with the aim to treat, diagnose, cure, mitigate, or prevent disease or other conditions is considered as a medical device.

---

4   There are, nonetheless, historical anomalies such as the GDPR, where the regulator took a rights-based approach to data management and data protection. Even though there are guidelines and requirements to perform risk assessments, the underlying idea is to clearly state and operationalise the citizens' rights in relation to their data.

5   Johanna Chamberlain, 'The risk-based approach of the European Union's proposed artificial intelligence regulation: Some comments from a tort law perspective' (2023) 14(1) European Journal of Risk Regulation 1-13 <https://doi.org/10.1017/err.2022.38> accessed 12 August 2023.

6   Tambiama Madiega, 'Artificial intelligence liability directive' (European Parliamentary Research Service, 2023) <https://www.europarl.europa.eu/RegData/etudes/BRIE/2023/739342/EPRS_BRI(2023)739342_EN.pdf> accessed 21 December 2023.

7   Ibid, 3.

8   See Alison Cave et al, 'Big Data – How to Realize the Promise' (2020) 107 Clinical Pharmacology & Therapeutics 753 <https://ascpt.onlinelibrary.wiley.com/doi/10.1002/cpt.1736> accessed 16 November 2023.

9   See Commission, 'White Paper on Artificial Intelligence – A European approach to excellence and trust' COM (20) 65 final.

10  Nakul Aggarwal et al, 'Artificial Intelligence in Healthcare' in Justin B Bullock et al (eds), *The Oxford Handbook of AI Governance* (1st edn, Oxford University Press 2022) <https://academic.oup.com/edited-volume/41989/chapter/355439553> accessed 14 August 2023.

11  Marjolein Timmers et al, 'How Do 66 European Institutional Review Boards Approve One Protocol for an International Prospective Observational Study on Traumatic Brain Injury? Experiences from the CENTER-TBI Study' (2020) 21 BMC Medical Ethics 36 <https://bmcmedethics.biomedcentral.com/articles/10.1186/s12910-020-00480-8> accessed 4 September 2023.

12  See Federal Food, Drug, and Cosmetic Act (FD&C Act), 21 CFR § 201(h) (2018).

13  See US Food and Drug Administration, 'Artificial Intelligence and Machine Learning in Software as a Medical Device' (*FDA*, 22 September 2021) <https://www.fda.gov/medical-devices/software-medical-device-samd/artificial-intelligence-and-machine-learning-software-medical-device> accessed 18 August 2023.

In this context, soft approaches to governance have been central and, in most cases, have spearheaded best practices in responsible research and innovation for cutting-edge technology. An important reason is that these governance approaches tend to be more agile than hard law while simultaneously requiring the direct involvement of the actors leading the path, partially addressing the pacing problem by closing the gap between regulatory efforts and technological progress.[14] This high degree of industry involvement is one of the main advantages of soft governance approaches as it streamlines processes and narrows the scope of specific measures or bodies. However, it can also water down the types of risk addressed in voluntary or self-regulating protocols as they usually gravitate around operational risks that are covered by existing standards and norms that validate, in turn, their own processes.

AI, nonetheless, is a type of general-purpose technology that has the potential to be revolutionary (or a type of technology that supports a fundamental transformation in the nature of economic production).[15] Before this possibility, self-governance and industry-wide soft governance efforts will play a crucial role in operationalising the guidelines and objectives defined collectively, bridging the distance between the government or regulator and the organisations developing AI. But because of AI's transformative potential, soft governance mechanisms must be coordinated and thoroughly scrutinised to ensure that they tackle the risks and threats relevant to the general public (beyond operational or corporate risks) and, ultimately, contribute to operationalise the underlying rights, values, and principles that constitute each society.

A rapid response mechanism to address the rising trend of AI-based research in healthcare has been to expand the scope of IRBs. However, IRBs were not originally formed to review AI-based projects. Considering this, and in the current regulatory context where AI is governed on a risk-based approach (with a pervasive focus on the risks that deployed systems entail), the legacy of IRBs' structures and processes, some of the concept definitions used (eg, regarding human subjects or what constitutes publicly available data), and a lack of resources and protocols to enforce their mandate could result in the challenges that IRBs face surpassing the governance opportunities they present.

## III. Main Challenges

According to the FDA's definition, IRBs are appropriately constituted groups that are:

> [...] formally designated to review and monitor biomedical research involving human subjects. In accordance with FDA regulations, an IRB has the authority to approve, require modifications in (to secure approval), or disapprove research. This group review serves an important role in the protection of the rights and welfare of human research subjects.
>
> The purpose of IRB review is to assure, both in advance and by periodic review, that appropriate steps are taken to protect the rights and welfare of humans participating as subjects in the research. To accomplish this purpose, IRBs use a group process to review research protocols and related materials (eg, informed consent documents and investigator brochures) to ensure protection of the rights and welfare of human subjects of research.[16]

As such, IRBs present different opportunities to govern the use and development of AI-enabled products and services in healthcare: on the one hand, the *raison d'être* of IRBs is to ensure that a set of different requirements, especially ethical guidelines, are observed in research projects, posing a distinctive opportunity to operationalise and embed ethically relevant discussions into the lifecycle of AI research projects. On the other hand, and closely related to this point, IRBs act upon different moments of the project's lifecycle or governance nodes, offering a unique longitudinal perspective over research projects. But, at the same time, and because of their almost exceptional status, IRBs face several challenges when assessing the risks entailed by AI-enabled research projects.

---

14  Gary E Marchant, Braden R Allenby and Joseph R Herkert (eds), *The Growing Gap Between Emerging Technologies and Legal-Ethical Oversight: The Pacing Problem*, vol 7 (Springer Netherlands 2011) <https://link.springer.com/10.1007/978-94-007-1356-7> accessed 12 September 2023.

15  Ben Garfinkel, 'The Impact of Artificial Intelligence: A Historical Perspective' in Justin B Bullock et al (eds), *The Oxford Handbook of AI Governance* (1st edn, Oxford University Press 2022) <https://academic.oup.com/edited-volume/41989/chapter/386766686> accessed 12 September 2023.

16  See US Food and Drug Administration, 'Institutional Review Boards Frequently Asked Questions: Guidance for Institutional Review Boards and Clinical Investigators' (FDA, January 1998) <https://www.fda.gov/regulatory-information/search-fda-guidance-documents/institutional-review-boards-frequently-asked-questions> accessed 4 September 2023.

In December 2022, the Ada Lovelace Institute published a report exploring the role of IRBs in evaluating AI-enabled research from an ethical standpoint to identify the main challenges these boards face. To do so, they conducted a literature review on the challenges faced by IRBs as well as several workshops with members of IRBs and researchers working on AI ethics.[17] From their findings, and in the context of this article, three challenges stood out:

– *Consistency:* An increasing interest in AI intersects with the nature of research projects in the domain of healthcare which are, in many cases, the product of the efforts from different actors and, sometimes, public-private partnerships. This is likely to trigger multiple evaluations of the same research project, highlighting in turn some of the challenges of governance and consistency between the standards observed among different IRBs.[18] This is not only problematic but could also lead to the 'gamification' of ethical reviews, exploiting the discrepancies among IRBs to the benefit of the promoter by selecting the better-suiting committee according to the project. Furthermore, private organisations are more and more well-positioned to conduct research exclusively overseen by their own 'advisory IRB', circumventing traditional review processes.
– *Exhaustiveness:* Many of the harms entailed by AI-enabled systems are only observable after the system has been deployed into the real world. However, IRBs are de facto tasked with assessing AI-enabled research projects *before* they are finalised, making it virtually impossible to predict some of the effects of the resulting product. Moreover, certain types of harms (spanning from bias to misuses of AI, but also in terms of the broader impacts of AI research in society) are hard to predict by a single committee.[19] The lack of tools to predict some of these harms and protocols to address wider societal concerns increases the chance of involuntary ethics washing, stressing the need to develop and coordinate thorough protocols.
– *Process transparency:* Private IRBs are closer to the ground truth, for they usually have privileged information about the research. The privilege is protected by trade secrets and NDAs, but results in a lack of transparency with respect to their processes that raises concerns from the research community and other stakeholders, ultimately putting at risk the efficacy of public-private research partnerships.[20] However, they serve on an advisory role, leaving the final decision to researchers or project managers. In this sense, a lack of transparency perpetuates the knowledge gap between research promoters and evaluators and fosters an information asymmetry between independent evaluators and research promoters.

But other than these, IRBs must also face:
– *Capacity Asymmetry:* Some private companies have the means to privately develop systems or services relying on powerful AI models that would qualify as medical products. These are often developed using publicly available data or data that is not necessarily private in a strict medical sense. The impact that these can have on society is tangible but given the limited mandate of IRBs and the lack of process transparency by private organisations, their R&D efforts are rarely overseen beyond a company-owned ethics review committee.[21] Ultimately, this poses a direct threat to the credibility and effectiveness of IRBs as a mechanism to govern R&D.

Addressing these challenges is crucial to ensure IRBs remain effective in governing the safe development of AI-based medical products and ability to provide a unique governance entry point at the R&D stages of AI.

## I. Harnessing R&D: A Deeper Look into the Context

AI R&D has boomed over the last decade, attracting talent and capital from all over the world. This boom has led to a substantial growth of the AI market and, therefore, the potential impact of AI over everyone's lives. To address this increased potential impact, several AI-related bills have been passed, going from 1 in 2016 to 37 in 2022 across 127 countries.[22]

---

17 Ada Lovelace Institute, 'Looking before we leap: Ethical review processes for AI and data science research' (*Ada Lovelace Institute*, 13 December 2022) <https://www.adalovelaceinstitute.org/report/looking-before-we-leap/> accessed 12 September 2023.
18 Ibid, 7.
19 Ibid.
20 Ibid, 7-8.
21 Phoebe Friesen et al, 'Governing AI-Driven Health Research: Are IRBs Up to the Task?' (2021) 43 Ethics & Human Research 35 <https://onlinelibrary.wiley.com/doi/10.1002/eahr.500085> accessed 18 July 2023.

However, AI R&D remains an under-regulated field, largely due to the rapid progress the AI has experienced in the last years, which clashes with longer timelines for hard regulation. This is problematic because some of the risks inherent to AI broadly, but even more so to cutting-edge AI systems, can be more easily redressed at the earlier research and development stages via safety and trustworthiness by design strategies. But without legally binding requirements, the effects of leaving R&D governance to market interests are simply unknown.

For example, the interest on the benefits and risks of GPT technology in chatbots applied to medicine,[23] or more ambitiously on the paradigm shift that foundation models can imbue into medicine and healthcare[24] is prescient. Yet the opacity throughout the research process and the corporate interests affecting the decision on which product to develop are not transparent, especially given the private nature of the actors leading the technological progress.

Google, Microsoft, Anthropic, and OpenAI announced the constitution of the 'Frontier Model Forum' to

> promote the safe and responsible development of frontier AI systems: advancing AI safety research, identifying best practices and standards, and facilitating information sharing among policymakers and industry.[25]

And while this announcement shows signs of some responsible research practices, it also strengthens the case for more blunt action upon the R&D stages of AI systems.

More concretely, IRBs are well positioned to affect a critical decision-point within the early stages of AI's lifecycle: the step from research to development. On the one hand, scientific research is intended to push the conceptual boundaries of the field while, on the other hand, development usually consists of leveraging said progress into concrete products or services with a target audience. This difference is often times blurry, yet the step from a strictly scientific pursuit to a well-directed effort to develop a product with a real-world application is not only possible after a large set of actors make concrete decisions that are strongly political in nature (in the sense that they ultimately determine how the public is affected as well as the nature of the benefit they afford) but is, perhaps, one of the key dividing moments in an AI-product lifecycle.

This division is significant, for even though there are some risks tied to general research[26] (eg, in developing new algorithms or expanding capabilities, particularly for larger models), material risks to society are ultimately the byproduct of the decisions made to develop a given technology into a product or service. As such, as it is the case with product regulation, risk-based approaches to regulation and governance target, by definition, existing products in each jurisdiction to prevent and protect consumers or users from a set of risks.

In this regard, considering certain aspects (or failing to do so) during the transition step from research to development and project definition determine some of the risks that the end-of-the-line product shall pose to society. As such, this is a critical decision point or governance node in the project lifecycle of an AI-based product. This is important because the development of an AI-based product or service does not follow the same lifecycle as other traditional products and, therefore, exempting R&D from independent oversight or legally binding requirements can lead up to inefficiencies in the resulting governance framework.

IRBs occupy a privileged position within the project lifecycle. Namely, IRBs oversee research projects regardless of the promoter – that is, they are the universal and independent gatekeepers – and do so from an early stage of the project's lifecycle. This confers IRBs' unique governance role in the AI in healthcare landscape. However, in the current context where the mechanisms to embody principles into practices are still under construction, IRBs are exposed and in need for protocols and material means to address the knowledge gap between the promoters and the evaluators.

---

22  Nestor Maslej et al, 'The AI Index 2023 Annual Report' (AI Index Steering Committee, Institute for Human-Centered AI, Stanford University, April 2023) <https://aiindex.stanford.edu/report/> accessed 18 July 2023.

23  Peter Lee, Sebastien Bubeck and Joseph Petro, 'Benefits, Limits, and Risks of GPT-4 as an AI Chatbot for Medicine' (2023) 388 New England Journal of Medicine 1233 <http://www.nejm.org/doi/10.1056/NEJMsr2214184> accessed 26 August 2023.

24  Michael Moor et al, 'Foundation Models for Generalist Medical Artificial Intelligence' (2023) 616 Nature 259 <https://www.nature.com/articles/s41586-023-05881-4> accessed 26 August 2023.

25  Verbatim from the press release by OpenAI. See OpenAI, 'Frontier Model Forum' (*OpenAI*, 18 August 2023) <https://openai.com/blog/frontier-model-forum#OpenAI> accessed 26 August 2023.

26  For Frontier Models, the emergent capabilities of which cannot be predicted before training and could lead to a change of paradigm, but also related to broader notions of safety, control, or privacy see Rishi Bommasani et al, 'On the Opportunities and Risks of Foundation Models' <https://arxiv.org/abs/2108.07258> accessed 26 August 2023.

The universality of IRBs, however, hints at the potentially disparate impact that longer review processes can have on smaller and medium-sized organisations doing research. Given the nature, composition, and workload of IRBs, review processes can take weeks or months, whereas research at the forefront of AI and, even more so, in the start-up context usually takes weeks, if not days.[27] Thus, the greater impact that compulsory review processes could have on start-ups and smaller and medium enterprises (in comparison with less money-constrained organisations) should be considered when drafting best practices and procedures for IRBs. This would not only ensure excellence in innovation for all parties involved but would also contribute to reducing power concentration by ensuring competitiveness among all types of organisation.

## II. Governance Levers and Sources of Authority

Governance levers include the different paths or mechanisms by virtue of which effective governance is induced – which are, in turn, closely tied to the source of authority on which they stand. For AI-enabled medical products, we can distinguish at least four governance levers and their respective sources of authority:

1. *Monitoring:* This lever does not refer to the activity conducted by IRBs on specific projects, but to the effort by the central authority to oversee that the practices followed by IRBs are compliant with their statutes and underlying laws, and that such practices address the risks that AI-enabled medical products pose. In the US context, monitoring is ultimately conducted by the FDA, the authority that has the power to declare and cease IRBs.
2. *Harmonising:* In a similar line, the central agencies like the FDA in the US have the role to harmonise the efforts put forward by different IRBs, making sure that all review boards observe standards equally, reducing the probability of promoters gamifying the review process.[28] On the other hand, if national or State laws generate asymmetries, harmonizsng efforts are intended to address them.
3. *Observing Norms and Standards:* Even though this lever consists of IRBs being familiar with relevant norms and standards and making sure researchers observe them, the underlying authority of this lever comes from the international bodies that develop best practices and desirable behaviours into norms and standards.
4. *Providing Training:* By means of disseminating the desired outcomes and standards across researchers and stakeholders, both the central authority and the different organisms or IRBs set the bar and ethical guidelines, which ultimately percolates into the researchers' best practices.

## I. Governance Nodes

Governance nodes are the different decision points throughout a project's lifecycle, in which a stakeholder can effect change and/or steer the stream of work, altering the behaviour of other actors and, ultimately, the end-result. IRBs, via different governance levers, and by virtue of their processes and protocols, have access to different moments of a project's lifecycle and can therefore shape the end-result with their decisions.

For AI-enabled projects, one of the most critical governance nodes is the moment in which general research is crystallised into a project proposal to develop a specific product. This draws the line between explorative research and concrete, market-facing development efforts, bringing in turn a unique opportunity to land principles into concrete red lines and practices to be observed during the following research.

---

27 Adam Molnar, David Stanley and Davide Valeriani, 'Neurotechnology, Stakeholders, and Neuroethics: Real Decisions and Trade-Offs from an Insider's Perspective' in Veljko Dubljević and Allen Coin (eds), *Policy, Identity, and Neurotechnology* (Springer International Publishing 2023) <https://link.springer.com/10.1007/978-3-031-26801-4_15> accessed 25 October 2023.

28 Organisms like the International Conference on Harmonisation (ICH) produce guidelines for global pharmaceutical development (which apply to Europe, USA, Canada, Japan, and Switzerland). These guidelines are intended to reduce duplication of clinical trials and improve the efficiency of assessment processes. Creating a discussion group within ICH or finding a way to display a similar approach to AI-based medical products could facilitate the safe circulation of AI-based medical products by strengthening the harmonisation governance lever.

In the report developed by The Ada Lovelace institute, a multi-stage ethics review process is proposed not only to break down the complexity and burden of the review process, but also to make justice to the governance nodes that appear throughout a given project's lifecycle, from the first draft designing the research protocol to post-publication reviews to understand how the evaluations made by the IRB have altered the results.[29] On the other hand, the transition from 'locked' models (or those models that, once developed, are not modified) to continuous learning approaches to AI and periodic modifications justifies this break down to streamline updates.

In this sense, the recommendation is to clearly determine the governance nodes in which an AI-enabled product or service is to be evaluated by the relevant IRB, including:

1. *Background:* In which the need for an AI-enabled solution affecting or relying on the use of real health data is justified and linked to theoretical results deriving from fundamental research.
2. *Product Development:* In which not only the scope of the product, but also the performance conditions or ethical considerations under which its release will be halted (eg, unequal performance across protected groups, or unacceptable failure modes).
3. *Results Publication:* In which relevant ethical considerations (both in terms of safety and possible harms to society) are systematically evaluated and included in the final report and/or product description.
4. *Relevant Modifications:* In which the processes to re-evaluate and allow future modifications, fine-tuning, or continuous learning is addressed.

These nodes apply to any AI-based medical product, yet some private organisations are capable of developing services and systems that would in practice qualify as medical products while circumventing traditional ethical review processes. These, even when overseen by company-owned ethical review committees, pose a structural threat to the trustworthiness of IRBs to effectively govern the R&D of AI in healthcare. As such, regulatory agencies should revisit foundational texts like the Common Rule, where critical terms used to demarcate the projects that need an independent ethical review are defined, with the aim of incorporating some of these private efforts. Doing so is particularly important considering the weight that 'publicly available' data is playing in the development of privately owned AI-based services, which can ultimately have a direct impact on the population and that, in many cases, make use of medical or medical-adjacent data – drawing again the line between general wellness products and medical devices.

## I. Target Behaviours and Actors

One of the challenges of IRBs as governance mechanisms for R&D is that we do not speak of a single actor, but of a network of independent boards. In this sense, governance comprises both public or private and individual or collective actors, and includes hard governance or regulation, industry-wide self-governance, and company self-governance.[30]

Central authorities grant the status, but the power is independently wielded by each board. Due to their proximity with R&D, boards and members have a direct impact on the final products delivered into the market from early stages. But to effectively activate the different governance levers, IRBs must be equipped to:

1. *Coordinate:* Mainly to avoid research promoters gamifying the review process. Even though IRBs are, by definition, independent, individual boards should be equipped and with universal basic means to ensure the preparedness of any board to address AI-based products. This is the instrumental to prevent IRBs becoming the enabler of ethics-washing efforts.
2. *Seek Coherence:* Through the coordination efforts mentioned above, the need for coherence with a set of foundational principles arises. In this sense, overarching bodies should clearly provide the baseline principles on which specific protocols operate, and require the inclusion of ethics experts, product engineers, or tech lawyers among others in IRBs to ensure that protocols are implemented observing such principles. These need not to be original or new but establish the foundation of any activity conducted by IRBs.[31]

---

29   Ada Lovelace Institute (n 17).

30   See Graeme Auld et al, 'Governing AI through Ethical Standards: Learning from the Experiences of Other Private Governance Initiatives' (2022) 29 Journal of European Public Policy 1822 <https://www.tandfonline.com/doi/full/10.1080/13501763.2022.2099449> accessed 25 October 2023.

3. *Seek Consistency:* On the one hand, pursuing internal consistency contributes to effective evaluations throughout the different governance nodes (eg, between the project definition stage and a subsequent review). Decisions should be traceable and expandable throughout the different governance nodes, and the nature of these should respond to each stage of the project (ie, listing potential risks in the background or first draft of the project, versus voicing concrete ethical concerns in the later stages). On the other hand, protocols and principles should be derived from a well-curated list of internationally relevant standards and best practices, both to ensure interoperability, international compatibility, and prevent an unnecessary *ad hoc* and heterogeneous maze of resources used by different boards across a given jurisdiction.

4. *Limit the Scope:* On the one hand, to ensure that the approval of a given product does not lead to the *de facto* approval of any subsequent modification of such product. On the other hand, to ensure that the IRB model is not stretched, as some point it already is.[32] By delimiting the mandate and the capabilities required to execute it for IRBs, upcoming needs will be more easily identified and potentially addressed.

   While IRBs have the capacity to withhold approval based on ethical considerations, it would be advisable to explicitly state their capacity to request changes in the fundamental research proposal (for example, asking for a less-invasive approach or technique to develop a ML model), which would be consistent with the mandate by the FDA of IRBs in the US to 'require modifications (to secure approval).'[33]

5. *Seek Transparency:*[34] Even though transparency is mainly instrumental (that is, it does not stand as a value in itself but as a means to an end), ensuring individual boards seek transparency is a means to first, reduce coordination costs among different boards and second, close the knowledge gap between project promoters and evaluators. In this case, transparency does not aim to facilitate algorithm explainability *per se* (that is, a better technical comprehension by a lay or non-involved audience) but rather capacity building and effective evaluation throughout IRBs.

## I. Final Discussion

IRBs occupy a unique position as governance mechanisms for AI-based medical products, for even though they exist in a regulatory context based on the risks posed by AI, they are conceived to voice ethical concerns during or even before research projects are executed. And while this could well mean that IRBs' mandate is being stretched beyond what these boards were initially designed to achieve, it can also be understood as an anomaly that presents four unique opportunities, with their respective risks and recommendations.

First, IRBs offer an opportunity to embed ethical principles into material governance protocols from the earliest stages of research. Nevertheless, and given the work required in terms of coherent guidelines and processes, review processes could lead to ethics washing if members of each one of the individual boards are not adequately equipped or surrounding norms and regulation lack teeth. Moreover, this proposal could also encourage the gamification of the review process, especially considering weak coordination among the nodes constituting the IRB network. To prevent such gamification, regulators must ensure harmonisation by agreeing on common baselines and providing active support to different boards.

Second, IRBs offer a chance to prove the increase in quality derive from overseeing R&D in AI and, potentially, to propagate the relevance of IRBs for AI R&D beyond the domain of healthcare. However, and given the nature of the review process, key stakeholders (mainly larger private organisations) could potentially find ways of circumventing or deferring IRBs' oversight, creating greater asymmetries among dif-ferent types of organisations, and reducing the impact of this governance proposal.

---

31 They could be inspired, for example, by the US Whitehouse's Blueprint for an AI Bill of Rights, which includes algorithmic discrimination protections, data privacy, notice and explanation, and human alternatives, consideration and fallback. See White House Office of Science and Technology, 'Blueprint For an AI Bill of Rights: Making Automated Systems Work for the American People' (October 2022) <https://www.whitehouse.gov/ostp/ai-bill-of-rights/> accessed 13 September 2023.

32 Phoebe Friesen et al, 'Governing AI-Driven Health Research: Are IRBs Up to the Task?' (2021) 43 Ethics & Human Research 35 <https://onlinelibrary.wiley.com/doi/10.1002/eahr.500085> accessed 18 July 2023.

33 See US Food and Drug Administration, 'Institutional Review Boards Frequently Asked Questions' (n 16).

34 For a more extensive discussion of transparency's role in AI, see Stefan Larsson and Fredrik Heintz, 'Transparency in Artificial Intelligence' (2020) 9 Internet Policy Review <https://policyreview.info/node/1469> accessed 20 August 2023.

Third, IRBs can be a way of reducing overall governance costs by legitimising themselves as standposts of R&D best practices. However, this could end up stretching even further the scope and mandate of IRBs, rendering them either as mere tick-boxes in an exercise of ethics washing or, at the other side of the spectrum, as the wardens of R&D excellence. To prevent this, the scope of IRBs should be clearly defined and backed up by a pool of resources and talent in accordance with the state of the art of AI.

Fourth and last, IRBs can tackle the pacing problem by bridging the gap between private organisations developing cutting-edge technology and public and/or independent boards reviewing their research efforts. Moreover, the lessons learned could be aggregated and used as input to develop codes of conduct and governance best practices to foster interoperability and global coordination. This, however, faces two risks.

On the one hand, the inherent asymmetry of power and interests between private actors and IRBs leaves most of the agency in the hands of the first, with limited incentives and unclear benefits. To address this, regulators should foster process transparency by establishing collaboration frameworks where key companies developing cutting-edge AI technology and IRBs could interact. This would practically act as a sandbox, where AI companies would benefit from the input of external and independent experts and IRBs would ensure their relevance as technology advances.

On the other hand, smaller and medium sized companies subject to the longer review processes of IRBs could see their capacity to innovate and to compete with larger organisations hampered. This would cause an asymmetry, given that larger organisations can usually absorb the cost of extended timelines more easily, disincentivising innovation among the smaller companies and increasing power concentration among the bigger ones. To avoid this, and to ensure competitiveness, best practices and procedures for IRBs should account for the potential disparate impact that extended timelines could have, establishing clear and defined timelines to include them in research planning and agile response mechanisms to address time-sensitive reviews.

Establishing a clear framework to govern research, development, and other key nodes in the life cycle of AI-based medical products through IRB oversight will require seizing the opportunities discussed above while addressing the risks discussed. Doing so will not only contribute to the governance map with a unique pathway to address issues related to the research and development phases of AI-based medical products but will also contribute towards developing trust and excellence in research, taking the quality of the next generation of AI products one step further.